# Proof that quantum mechanics is internally inconsistent on antihydrogen

G. Van Hooydonk, Ghent University, Faculty of Sciences, Krijgslaan 281, B-9000 Ghent (Belgium)

**Abstract.** Quantum mechanics (QM) theoretically forbids natural H$\underline{\text{H}}$ interactions because of annihilation in the Dirac sense. But in practice, ab initio QM relies explicitly on H$\underline{\text{H}}$ attraction, which it theoretically forbids, to arrive at attractive forces in the molecular hydrogen cation, given away by the cusp in its PEC (potential energy curve) at exactly 1,06 Å. This internal inconsistency in QM is easily removed by lifting its irrational ban on natural $\underline{\text{H}}$.
Pacs: 34.10.+x, 34.90.+q, 36.10.-k

### Introduction

With the premises of QM, H$\underline{\text{H}}$ is an *exotic* system. QM cannot accept signatures for natural $\underline{\text{H}}$-states in the line spectrum of atomic hydrogen [1a] as well as in the band spectrum of molecular hydrogen [1b]. These signatures contradict the premises of QM as well as *its ban on natural $\underline{\text{H}}$* and on antimatter in general, a great problem for cosmology [2]. Despite this generally approved ban, interest in the properties of $\underline{\text{H}}$ remains for various reasons [3], as H (e$^-$p$^+$) and $\underline{\text{H}}$ (e$^+$p$^-$) only differ by charge-antisymmetry. The ban implies that HH interactions use only *proton-proton repulsion* $+1/R$ to explain system stability, while H$\underline{\text{H}}$ interactions with a *proton-antiproton attraction* $-1/R$ are forbidden. But vetoing H$\underline{\text{H}}$ attraction implies that H$\underline{\text{H}}$ annihilation must come to the rescue to assure QM remains consistent. But also H$\underline{\text{H}}$ annihilation prohibits signatures for natural $\underline{\text{H}}$, even when these are clearly visible in the spectra [1]. This impasse probably results from *something very elementary* that went wrong very early *with the theoretical perception of H$\underline{\text{H}}$ and of matter-antimatter interactions in QM*. This is evident from contradicting *ab initio* QM H$\underline{\text{H}}$ PECs [4], since only the first H$\underline{\text{H}}$ PECs showed cusps, in line with the cusp for similar 4-unit charge system H$_2$ [5]. A cusp would not only affect the H$\underline{\text{H}}$ annihilation cross-section but also point to a transition from *repulsive $+1/R$ to attractive $-1/R$*, forbidden in QM [4]. The cusps being annoying for QM, the search for cusp-less QM H$\underline{\text{H}}$ PECs started, inspired if not biased by the cusp-less PEC for annihilative channel Ps+Pn [4] and clearly inspired by [3]. *Despite its mathematical rigor, we easily prove that QM is deceptive on H$\underline{\text{H}}$* [1c]. We prove this for *the molecular hydrogen cation, with a complexity in between that of atomic and molecular hydrogen*. With charge-antisymmetry, this cation would have *2 Coulomb quantum states* $\pm 1$, one for *hydrogenic cation* HH$^+$; the other for *antihydrogenic cation* $\underline{\text{H}}$H$^+$ [1c]. To make sense, *both cation states $\pm 1$ must show in the cation's ab initio QM PEC*, which is exactly what we will prove analytically.

### QM bonding in the molecular hydrogen cation: repulsion $+e^2/r_{AB}$ and cusp problem

Different symmetries for bonding and anti-bonding states of the molecular hydrogen cation are given away by their PECs. Asymmetry may derive from positional coordinates but it is certain that antisymmetry derives from intra-atomic charge inversion [1], not considered at the time of the Pauling-Wilson (PW) QM method [6]. Their method is not biased by annihilation, if a H$\underline{\text{H}}$ interaction would appear. Their solutions for the secular equation, obtained *in tempore non suspecto.*, are symmetric $W_S$ and antisymmetric $W_A$

$$W_S = W_H + e^2/a_0 D + (J+K)/(1+S) \tag{1a}$$
$$W_A = W_H + e^2/a_0 D + (J-K/(1-S)) \tag{1b}$$

where separation $r_{AB}$ for Coulomb nucleon repulsion $+e^2/r_{AB}$ has is scaled with $a_0$ using standard notation

$$D = r_{AB}/a_0 \tag{1c}$$

$W_H = -\tfrac{1}{2}e^2/a_0$ is the eigenvalue for atom H, J and K are 1- and 2-center integrals and S represents the lack of orthogonality of atomic functions (see below). With (1a,b), cation stability relies on proton-proton repulsion $+e^2/r_{AB}$, the only *classical Coulomb term* in (1a,b). The same term appears in QM for molecular hydrogen H$_2$ [5,6], suggesting that also here proton-proton repulsion leads to stable natural species H$_2$. *Positive* $+e^2/r_{AB}$ for HH$^+$ relates to *negative* $W_H$, the leading terms of opposite sign in ab initio QM (1a,b). Therefore, scaling $W_S$ and $W_A$ gives a scaled *repulsive Coulomb relation* of form $-1+2/D$ for (1a,b).

However, the exclusivity of *proton-proton repulsion* is refuted by the cusp in the HH$^+$ PEC, needed to explain the cation's stability [6] and suggests equilibrium between repulsion and attraction.

*QM repulsive branch $-1+2/D$ in (1a,b) proves that only attractive branch $+1-2/D$ can cope with this QM repulsion. If and only if attraction $+1-2/D$, forbidden by QM, really exists, a cusp automatically appears when $+1-2/D=-1+2/D$. With asymptotes $\pm 1$, the Coulomb cusp appears exactly at D=2, i.e. at $r_{AB}=2a_0=1,06$ Å. Using asymptotes $\pm 1$ secures that cusp formation is understood classically with Coulomb's law. A disadvantage of a symmetrical solution for perfectly balanced attraction and repulsion is that the resulting PEC always vanishes exactly, since*

$$(+1-2/D)+(-1+2/D) \equiv 0 \tag{1d}$$

*A non-zero result is the equivalent of an observable PEC centered along the zero asymptote but this can only be obtained with asymmetry in the Coulomb interactions.* The greatest difficulty with (1d) that it is zero for perfect symmetry, which means that it could prove very difficult to retrace (1d) in ab initio QM. But interest in Coulomb models derived from *the observed cusp for the cation, exactly at generic Coulomb value $r_{AB}=1,06$ Å* [6] (see below).

**Coulomb view on bonding in the molecular hydrogen cation**

QM could be right to rely exclusively on proton-proton repulsion, if it were not for the cusp and its position at D=2. Charge-conjugated Hamiltonians [1c] for the cation give Coulomb pair $\pm e^2/r_{AB}$, for repulsion and attraction. The algebraic Hamiltonian ($\mathbf{H_+}$ for hydrogenic HH$^+$ and $\mathbf{H_-}$ for antihydrogenic $\underline{H}H^+$) obeys [1c]

$$\mathbf{H_\pm = H_0 \pm \Delta H = H_0 \pm (-e^2/r_B + e^2/r_{AB})} \tag{2a}$$

with $W_H$ as eigenvalue for $\mathbf{H_0}$. Obviously, QM only accepts $\mathbf{H_+}$ and forbids $\mathbf{H_-}$. Algebraic perturbation $\pm\mathbf{\Delta H}$ contains formally[1] conjugated pair $\pm e^2/r_{AB}$ for *nucleons*: $+\mathbf{\Delta H}$ for HH$^+$ gives *repulsive* $+2/D$ for the nucleon part of (2a) as it appears in (1a,b), whereas $-\mathbf{\Delta H}$ for $\underline{H}H^+$ gives *attractive* $-2/D$, forbidden in QM, which only tolerates $+2/D$ for the nucleons. The absolute value of the perturbation depends on $r_B$ and $r_{AB}$ (see footnote 1). But because of the formal importance attached by QM to the nucleon part in (1a,b), we concentrate on term $\pm 2/D$. We rely on the PW ab initio QM approach [6] to deal with the effects of positional coordinates on lepton-nucleon term $\pm e^2/r_B$ in (2a) and the role of integrals J and K is acknowledged further below. Whatever the further interpretation of (2a), its direct implication is that it indeed inverts the QM nucleon repulsion $+2/D$ in (1a,b) into nucleon attraction $-2/D$, *absolutely forbidden in QM* (see Introduction)

Having said this, we must only retrieve a nucleonic attraction for the cation in ab initio QM to prove our thesis and to validate, in general, the reality of the signatures we already found for natural H [1]. As argued in [1c], the *conventional solution is avoiding the problem of finding attraction-2/D, needed with (2a). This solution[2] uses circular reasoning by saying that (2a) is wrong because it is wrong: it allows proton-antiproton attraction, which is forbidden by QM because of annihilation (see Introduction).*

*A better scientific solution is to look for attractive branch +1-2/D, by verifying it is not hidden somewhere in QM, although this also seems very unlikely at first sight when looking at QM result (1a,b) and at the premises of QM. But, if this branch were really hidden in QM, the QM machinery would prove deceptive on H$\underline{H}$, if not internally inconsistent completely.*

This bold far-reaching hypothesis is much easier to verify than in [1c] for H$_2$, since with the PW method for the cation all relatively simple integrals are available analytically [6]. State HH$^+$, with asymptote $-1$ and repulsion $+2/D$ is, by definition, visible in QM result (1a,b), whereas state $\underline{H}H^+$, with asymptote $+1$ and attraction $-2/D$ *is invisible in QM and must be so to respect its own premises.*

Combining all Coulomb information gives 2 branches

$$W_\pm(D) = \pm(1-2/D) \tag{2b}$$

in line with (1d). Without *quantum states* $\pm 1$ (asymptotes), these simplify to

$$W^0_\pm(D) = \pm(-2/D) \tag{2c}$$

like the original Coulomb law and its built-in antisymmetry. Whatever the shape of the PEC, the different formulation for the same law has to do with cusp formation, since pair (2b) directly provides a cusp at exactly D=2, which is impossible with pair (2c) (see above).

**Retrieving forbidden $\underline{H}H^+$ attraction +1-2/D in the ab initio QM calculation**

To find out about attractive branch +1-2/D, we make advantage of all the qualities of the PW scheme [6]: it has *ab initio* status, all integrals are available analytically, it is transparent and *was written in tempore non suspecto*. Now, *finding +1-2/D in QM only requires 3 mathematically simple, almost trivial but physically important steps.*

(i) Scaling $W_S$ and $W_A$ in (1a,b) with half the Hartree ($+½e^2/a_0$) leads to

$$W_S/(+½e^2/a_0) = W'_S = -1 + 2/D + 2(J'+K')/(1+S) \tag{3a}$$

$$W_A/(+½e^2/a_0) = W'_A = -1 + 2/D + 2(J'-K')/(1-S) \tag{3b}$$

where J' and K' are scaled J and K in (1a,b). This exposure of repulsive branch $-1+2/D$ is *trivial* in QM.

(ii) Shifting these two scaled PECs towards the zero asymptote gives $W^0_S$ and $W^0_A$, which requires another *seemingly trivial* step, i.e. adding asymptote $+1$ in each equation (3) or

$$W^0_S = -1 + 2/D + 2(J'+K')/(1+S) +1 \tag{4a}$$

$$W^0_A = -1 + 2/D + 2(J'-K')/(1-S) +1 \tag{4b}$$

which formally and unequivocally leads to asymptote $+1$ for $\underline{H}H^+$, needed for (2b) to make sense. *Exactly and already at this trivial stage*, a first deception of QM appears. In fact, one is tempted to replace the asymptote difference $+1-1$ in (4) by zero, which is mathematically correct. But using zero wipes out the 2 Coulomb quantum states $\pm 1$, explicitly needed for Coulomb formula (2b) to explain cusp formation. Using only (2c), a zero asymptote Coulomb formulation would have left us *without cusp formation*. Since replacing 0 by $+1-1$ is a major achievement of quantum theory, one must be very cautious with these mathematically correct, *if not trivial*, replacements. In fact, after having found asymptote $+1$ for $\underline{H}H^+$ in (4a,b), we are left with finding attractive $-2/D$, the only missing link to connect ab initio QM with $\underline{H}H^+$ *attraction* introduced with (2a).

---

[1] The relative contribution of $r_B$ and $r_{AB}$ to the total perturbation (positional coordinates) can affect the influence of charge-inversion [7]. Here, we nevertheless use *nucleonic* $\pm 2/D$, since $2/D$ is the main variable also in QM [6].

[2] This explains but does not justify the reluctance of the establishment with signatures for natural $\underline{H}$ [1].

(iii) Before scaling to J' and K', 1 and 2-center integrals J and K are [6]

$$J = (e^2/a_0)(-1/D + e^{-2D}(1+1/D)) \quad (5a)$$
$$K = -(e^2/a_0)e^{-D}(1+D) \quad (5b)$$

We remind that hydrogenic STOs functions $u_{1s(A)}$ and $u_{1s(B)}$ are used by Pauling and Wilson [6] and that the 2-center functions are symmetric and antisymmetric in positional coordinates only

$$\psi_S = (u_{1s(A)} + u_{1s(B)})/(2+2S^2)^{1/2} \quad (6a)$$
$$\psi_A = (u_{1s(A)} - u_{1s(B)})/(2-2S^2)^{1/2} \quad (6b)$$

where S represents the lack of orthogonality of the atomic functions (see above).
As in (4), scaling J and K gives

$$J' = -2/D + 2e^{-2D}(1+1/D) \quad (7a)$$
$$K' = -2e^{-D}(1+D) \quad (7b)$$

This last step in one center integral J (7a), *not in* two-center K (7b), produces exactly the missing link $-2/D$, still required. Only, its real contribution to the total PEC is attenuated by $1/(1+S)$ because of the lack of orthogonality of the atomic functions, giving formally

$$-2/D \approx (-2/D)/(1+S) \quad (7c)$$

but this must not distract us from our main argument. In good first order approximation (S=0), the classical Coulomb branch $+1-2/D$ in (2b) is retrieved exactly in ab initio QM. Moreover, resonance or exchange integral K', seemingly absent in classical physics, refers to 2 asymptotes instead of only 1, just like S, since atomic functions $u_{1s(A)}$ and $u_{1s(B)}$ are both required for their evaluation. Despite the complexity of ab initio QM, we identified one by one all of the terms required for classical Coulomb scheme (2b), e.g. its four terms -1, +1, +2/D and -2/D, although the last term appears only in a good first order approximation (7c). The unambiguous conclusion is that the classical Coulomb PECs (2b,c), generated by algebraic Hamiltonian (2a), are, *seemingly unwillingly or unknowingly*, used almost identically in ab initio QM [6].

*Ab initio QM itself not only proves analytically that HH+ attraction -2/D appears in nature but also that this is essential to explain the stability of the molecular hydrogen cation. This is why QM is so deceptive on HH. First, conventional QM tries to persuade us theoretically that HH attraction is forbidden in nature since it leads to annihilation. But in practice, QM –secretly- uses the HH attraction it forbids to explain classically the stability of the molecular hydrogen cation.*

These results prove unambiguously that QM is inconsistent on HH [1c].

**Results and consequences**
*Ab initio QM versions of Coulomb PECs*
Since K<0, the QM PECs for cation with explicit quantum states +1-1 can now be rewritten as

$$W_{\pm S} = [-1 +2/D +2e^{-2D}(1+1/D)]_r + [1 -(2/D)/(1+S) +2e^{-D}(1+D)/(1+S)]_a \quad (8a)$$
$$W_{\pm A} = [-1 +2/D +2e^{-D}(e^{-D}(1+1/D)-(1+D))/(1-S)]_r + [1 -(2/D)/(1-S)]_a \quad (8b)$$

which, apart from irrelevant scale factor ½, are numerically identical with (4a,b). These are the QM versions of Coulomb recipe (2a,b), as *hidden* in the ab initio QM PW calculation [2]. All *positive, repulsive* terms for –1 or HH+, are between square brackets with subscript *r*, and all *negative attractive* terms for +1 or HH+ are between square brackets with subscript *a* [1c]. In (8a,b), the *classical Coulomb* branches $-1+2/D$ for HH+ and $+1-2/D$ for HH+ are perfectly in line with Coulomb expectations (2a,c), although $-2/D$ is attenuated by $1/(1+S)$, due to the lack of orthogonality of the atomic functions (see above).

Reminding (2), only the *generic Coulomb PECs* can predict an equally *generic cusp* when $W_\pm(D)=0$, which is at

$$D = 2 \text{ or } r_{AB} = 2a_0 = 1,06 \text{ Å} \quad (9)$$

Surprising or not, this is exactly the equilibrium separation for the molecular hydrogen cation [6].
Without asymptotes ±1, the PECs are mathematically identical with (8a,b) but *completely different in terms of physics*. We denote them with a zero superscript

$$W^0_{\pm S} = [+2/D +2e^{-2D}(1+1/D)]_r - [(2/D)/(1+S) - 2e^{-D}(1+D)/(1+S)]_a \quad (10a)$$
$$W^0_{\pm A} = [+2/D +2e^{-D}(e^{-D}(1+1/D) -(1+D))/(1-S)]_r - [(2/D)/(1-S)]_a \quad (10b)$$

*Symmetric solutions* for Coulomb repulsion and attraction of type *(rep+att)/2* are, by definition, always centered along the zero asymptote, whether asymptotes +1 and –1 are included or not. The small QM *asymmetry* for the forces is, essentially, due to *positional coordinates* [6,7]. But the *antisymmetry* of (2a), *generated by intra-atomic charge-inversion*, is not really showing but *hidden in the QM framework*, the main reason why QM is deceptive [1c]. The better *antisymmetric solutions* for Coulomb repulsion and attraction of type *(att – rep)/2*, where attractive forces start off at non-zero, positive asymptote (+1) are given elsewhere [1b,1c]. These are the so-called *ionic approximations*, whereas those of QM are *covalent approximations* [1c].

*Graphical illustrations*
Fig. 1a gives a plot of the branches in (8a) and (10a) and the PEC they generate for the bound state in function of D. It is clearly shown how the QM PEC for the bound state of the cation derives from adding 2 conjugated branches in (8a) and (10a). To illustrate the correspondence with the pure Coulomb view on bonding, we included PEC branches (2b) and (2c). This surprising result has, to the best of my knowledge, never been reported before. Fig. 1b is the plot with 1/D [1c] and is even more illustrative. At large D, the classical and QM view on the cation *coincide*, meaning that the 2 methods effectively start with the two Coulomb quantum states -1 for HH$^+$ and +1 for $\underline{H}$H$^+$, deriving explicitly from algebraic Hamiltonian (2a). We also readily verify how only the PECs *with the 2 quantum states ±1* lead to the generic cusp at D=2 we wanted by virtue of (2b). The observed equilibrium separation being at the generic Coulomb value D=2, the practical ab initio QM result at $D_{QM}$=2,51 [6] indicates that the PW approach *for deviations from pure Coulomb attraction and repulsion* can be improved, *since their final result does not respect the Coulomb cusp*.

The same plots show that the more classical Coulomb presentation (±2/D) in (2c) starts off at the zero asymptote at infinite D, which is the reason why they can never generate a cusp. In fine, the cusp in the observed PEC can only derive, *even in the practice of an ab initio QM approach*, from $\underline{H}$H$^+$ attraction in the classical Coulomb sense, *although this is theoretically forbidden in QM*. This contradiction survived for decades and denied, amongst others, the option to denote natural H$_2$ as H$\underline{H}$ instead [1c].

*Further consequences*

(i) Fig. 1a and 1b are conceptually useful as they show, at last, how intimately the 20$^{th}$ century QM approach to chemical bonding is related with the classical ionic Coulomb bonding ideas of the early 19$^{th}$ century [1c].

(ii) The *adiabatic Born-Oppenheimer approximation (BOA)* was used for QM H$\underline{H}$ PECs [4,6] (see Introduction). The BOA separates *the nucleon-nucleon repulsion* from lepton-nucleon and lepton-lepton interactions on order to solve the *leptonic* wave equation. Now, *nucleon-nucleon attraction*, absent in the BOA, is essential for cusp formation as illustrated in Fig. 1, which places question marks on the meaning of the BOA.

(iii) When the Coulomb gap between two states occupied by H and $\underline{H}$ respectively is also a quantum gap, natural H$\underline{H}$ oscillations become plausible, although these are forbidden also in QM. We proved elsewhere that, for these H$\underline{H}$ transitions, oscillation times of about 10$^{-15}$ sec are expected, whereas estimates with conventional physics give 10$^{+20}$ sec [7].

(iv) A close investigation of the interplay of discrete anti-symmetry by charge-inversion with asymmetries or pseudo-antisymmetry generated by positional coordinates (particle geometries) seems worthwhile [8]. A similar problem is looking for ways to separate the wave function using different coordinate systems, *while at the same time intra-atomic charge-inversion with its built-in antisymmetry is not considered.*

(v) As indicated in [1c], one may wonder indeed if Pauli spin antisymmetry is really required, when charge-antisymmetry is so readily available? The operators are identical, except for a scale factor of 2 [1c].

A full discussion of these few examples is, however, beyond the scope of the present work.

*Matter and antimatter in the Universe*

For cosmology, the matter-antimatter asymmetry of the Universe [2] seems like *a discussion about nothing*, since

$$0 = +1 - 1 \qquad (11)$$

is important for antimatter (antihydrogen) and appears *exactly but secretly* in *ab initio QM*. Using 0 is much like classical *Coulomb* physics, whereas quantum state notation +1-1 seems a great result of quantum theory. Even this is deceptive, since quantum states ±1 for any system, *and immediately appearing as soon as one tries to describe this system with a Coulomb model*, are exactly those of the Coulomb laws for attraction and repulsion, i.e. ±1/R.

**Conclusion**

Both the conventional *a priori ban* in QM on natural $\underline{H}$ (antimatter) and its interactions with H (matter) as well as the presumed annihilation of pair $\underline{H}$H are easily falsified. The irony is that the proof for this falsification rests with *QM* itself. QM proves *deceptive and even internally inconsistent on H$\underline{H}$* and will remain so unless its irrational ban on natural $\underline{H}$ is removed, like we suggested early in Einstein/Physics Year 2005 [1c]. Why QM succeeded in *wrong footing* so many for so long on *natural $\underline{H}$* is and will remain an open question, with *wrong* [1,3b] experiments on *artificial $\underline{H}$* like [3a], scheduled at CERN and GSI at considerable cost.

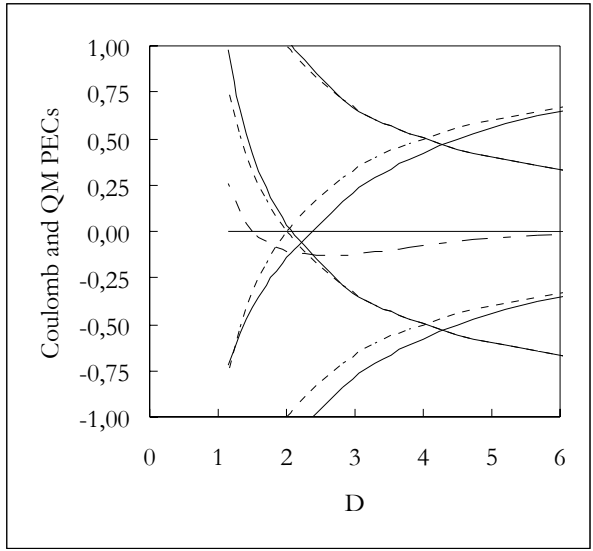 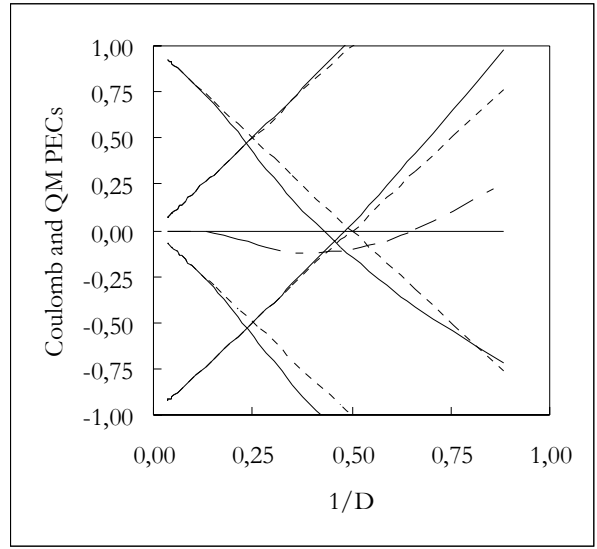

Fig. 1a Molecular hydrogen cation PECs versus D QM crossing (8a) and non crossing (10a) branches (full), Coulomb references (2b) and (2c) (dashes), 2 times PW PEC (mixed dashes), zero Coulomb PEC (1d) (full)

Fig. 1b Molecular hydrogen cation PECs versus 1/D same notation as in Fig. 1a